\documentclass{PoS}

\usepackage{amsmath,latexsym}
\usepackage{physics}
\usepackage{amsfonts}
\usepackage{color}
\newtheorem{problem}{Problem}

\title{The Very Basics of Higher-Spin Theory}

\ShortTitle{The Very Basics of Higher-Spin Theory}

\author{\speaker{Pan Kessel}\\
        Max Planck Institute for Gravitational Physics (Albert Einstein Institute)\\
        E-mail: \email{pan.kessel@aei.mpg.de}}

\abstract{These notes are based on two lectures given at the
Twelfth Modave Summer School in Mathematical Physics 2016. The Fronsdal equation and action for both Minkowski and (A)dS backgrounds are discussed in detail.}

\FullConference{XII Modave Summer School in Mathematical Physics\\
		12-16 September 2016\\
		Modave, Belgium}

\begin{document}

%----------------------------------------------------------------
\section{Introduction}
These notes are based on a two one-hour lectures given at the Twelfth Modave Summer School in Mathematical Physics 2016 in which I tried to give master and PhD students a good working knowledge of the very basics of higher-spin gauge theories. The free theory of fully symmetric massless higher-spin fields is discussed in detail. I tried to make the lectures as interactive as possible which is, of course, hard to reproduce in this write-up. Nevertheless, I hope to preserve at least the spirit of these lectures by adding various exercises and their solutions. The reader is strongly encouraged to work through them.

The material presented here is absolutely elementary and contains no original results. I hope that it provides a good start for studying more advanced and intermediate concepts of higher-spin theory.

There are a number of useful resources on this subject. Three references from which I particularly benefited while preparing the lectures are

\begin{itemize}
	\item Section 2 of \cite{Rahman:2015pzl} which also discusses in detail fermionic and massive higher-spin fields.
	\item Section 2 of \cite{Didenko:2014dwa} which in later chapters also provides a useful starting point to learn about more advanced aspects of higher-spin theories, in particular Vasiliev theory.
	\item Section 2 of \cite{Campoleoni:2010zq}: this reference also discusses the particularities of three-dimensional higher-spin theories.
\end{itemize}

\noindent For some parts of the discussion, I follow these references quite closely.

%----------------------------------------------------------------
\section{Fronsdal Equation}
\label{sec:FronsdalEquation}
Gauge fields are one of the most important building blocks of modern theoretical physics. For example, the standard model of particle physics contains various spin-$1$ gauge fields which lead to the electroweak and strong force. The most simple example of a spin-1 gauge field is given by electromagnetism. Maxwell's equation can be written in the following form
\begin{equation}
\label{eq:maxwellEquations}
\partial^\mu F_{\mu \nu} = 0 \,,
\end{equation}
where the spacetime indices are denoted by $\mu,\nu,\ldots \in \{0,1,\ldots,D-1\}$ and the field strength tensor is given in terms of the spin-$1$ field $A_\mu$ by
\begin{equation}
\label{eq:fieldStrength}
F_{\mu \nu} = \partial_\mu A_\nu - \partial_\nu A_\mu \,.
\end{equation} 
By inserting \eqref{eq:fieldStrength} in \eqref{eq:maxwellEquations}, one obtains
\begin{equation}
\label{eq:spin1EoM}
\Box A_\mu - \partial_\mu \partial^\sigma A_\sigma = 0 \,.
\end{equation} 
Obviously, the field strength \eqref{eq:fieldStrength} is invariant under the following spin-$1$ gauge transformation
\begin{equation}
\delta A_\mu(x)  = \partial_\mu \xi(x) \,,
\end{equation}
where $\xi$ is an arbitrary function of the spacetime coordinates $x^\mu$. As a result, also the equation of motion \eqref{eq:spin1EoM} is invariant under this gauge transformation. 

Spin-1 gauge fields can be used to describe all fundamental forces of nature but gravity. Note however that the standard model contains Yang-Mills gauge fields whose equations of motion are similar to the abelian Maxwell theory \eqref{eq:spin1EoM} but also contain additional interaction terms. Only when we neglect these interaction terms by restricting to terms linear in the gauge fields do we obtain free equations of motion of the form \eqref{eq:spin1EoM}. 

In the case of gravity, free equations of motion can be obtained from the vacuum Einstein equations
\begin{equation}
R_{\mu \nu} = 0
\end{equation}
by rewriting the metric as $g_{\mu \nu} = \eta_{\mu \nu} + h_{\mu \nu}$ and only keeping terms linear in $h_{\mu \nu}$, one then obtains
\begin{equation}
\label{eq:spin2EoM}
\Box h_{\mu \nu} - \partial_{(\mu} \partial^\sigma h_{\nu)\sigma} - \partial_{\mu} \partial_{\nu} h^\sigma{}_\sigma = 0 \,,
\end{equation}
which can be checked to be gauge invariant under the following spin-2 gauge transformation
\begin{equation}
\delta h_{\mu \nu} = \partial_{(\mu} \xi_{\nu)} \,.
\end{equation}
For a summary of our symmetrization conventions see Appendix \ref{app:conventions}.

Since general relativity and the standard model are described by spin-2 and spin-1 gauge fields respectively (in the sense we have explained above), it is therefore tempting to generalize the equations of motions \eqref{eq:spin1EoM} and \eqref{eq:spin2EoM} to arbitrary spin-$s$. A natural ansatz\footnote{At this stage, one might wonder why there are no terms involving double-traces $\phi_{\mu_1 \dots \mu_{s-4} \sigma \lambda}{}^{\sigma \lambda}$ and also higher traces. In Section \ref{sec:DOFofFronsdal}, we will impose double-tracelessness condition $\phi_{\mu_1 \dots \mu_{s-4} \sigma \lambda}{}^{\sigma \lambda}=0$ in order to show that the equations of motion propagate the correct degrees of freedom. Therefore, we do not include these terms in \eqref{eq:fronsdalEquation}.} for this is given by the Fronsdal equation \cite{Fronsdal:1978rb}
\begin{equation}
\label{eq:fronsdalEquation}
\boxed{F_{\mu_1 \dots \mu_s} = \Box \phi_{\mu_1 \dots \mu_s} - \partial_{(\mu_1} \partial^\sigma \phi_{\mu_2 \dots \mu_s) \sigma} + \partial_{(\mu_1} \partial_{\mu_2} \phi_{\mu_3 \dots \mu_s)\sigma}{}^{\sigma} = 0 }\,,
\end{equation}
where $F_{\mu_1 \dots \mu_s}$ is called the Fronsdal tensor. The gauge transformation of the Fronsdal field is given by
\begin{equation}
\delta \phi_{\mu_1 \dots \mu_s} = \partial_{(\mu_1} \xi_{\mu_2\dots\mu_s)} \,.
\end{equation}
However, one can easily show that the gauge variation of the Fronsdal tensor is proportional to the trace of the spin-$s$ gauge parameter
\begin{equation}
\delta F_{\mu_1 \dots \mu_s} = 3 \, \partial_{(\mu_1} \partial_{\mu_2} \partial_{\mu_3} \xi_{\mu_3 \dots \mu_s) \sigma}{}^{\sigma} \,.
\end{equation}

\begin{problem}
Calculate the gauge variation of the Fronsdal tensor $F_{\mu_1 \dots \mu_s}$.
\end{problem}

Therefore, in order to ensure that the Fronsdal equation is gauge invariant, we have to require\footnote{In principle, one could also attempt to impose a differential constraint on the gauge parameter. Let us consider $s=3$ for definiteness. In this case, the most general solution of $\partial_{\mu_1}\partial_{\mu_1}\partial_{\mu_1} \xi'=0$ is a polynomial of degree two. If one requires that the gauge parameter vanishes at infinity, the only solution is $\xi_{\sigma}{}^\sigma \equiv 0$. In the case of $s>3$, an analogous argument leads to the same conclusion.} that the gauge parameter is traceless
\begin{equation}
\label{eq:tracelessnessOfGaugeParameter}
\xi_{\mu_1 \dots \mu_{s-3} \sigma}{}^{\sigma} = 0 \,.
\end{equation}

%----------------------------------------------------------------
\section{Degrees of Freedom}
In the last section, we have presented a natural ansatz \eqref{eq:fronsdalEquation} for an equation of motion of a spin-$s$ field $\phi_{\mu_1 \dots \mu_s}$ by generalizing the Maxwell and linearized Einstein equations.  In the following, we will show that the Fronsdal field $\phi_{\mu_1 \dots \mu_s}$ indeed propagates the correct number of degrees of freedom of a massless spin-$s$ field in $D$-dimensional Minkowski spacetime which is given by
\begin{equation}
\label{eq:numOfDegOfFreed}
\binom{D-3+s}{s} - \binom{D-5+s}{s-2} \,.
\end{equation}

In order to derive this number, we first need to review the Wigner method for constructing unitary irreducible representations of the Poincar\'e algebra which we will briefly sketch in the following. The discussion will be slightly technical. \emph{Upon first reading, the reader might want to take \eqref{eq:numOfDegOfFreed} as given and skip to Section \ref{sec:DOFofFronsdal}.} 

For a detailed account of the Wigner method, we recommend Chapter 2 of Weinberg's book 'The Quantum Theory of Fields' \cite{Weinberg:1995mt} and lecture notes by Bekaert and Boulanger \cite{Bekaert:2006py} which also contain applications to higher-spin theory.

\subsection{Wigner Classification}
In quantum field theory, we associate particles with unitary irreducible representations of the Poincar\'e algebra
\begin{subequations}
\begin{align}
[M_{\mu\nu},M_{\rho \sigma}] &= i (\eta_{\mu \rho} M_{\nu \sigma} - \eta_{\nu \rho} M_{\mu \sigma} - \eta_{\mu \sigma} M_{\nu \rho} + \eta_{\nu \sigma} M_{\mu \rho}) \,, \\
[P_\mu, M_{\rho \sigma}] &= -i (\eta_{\mu \rho} P_{\sigma} - \eta_{\mu \sigma} P_{\rho}) \,, \\
[P_\mu, P_\nu] &= 0 \,.
\end{align}
\end{subequations}
These representations can be found using the Wigner method which can be roughly summarized as follows: the generators $P_\mu$ should be realized as self-adjoint, commuting operators and can therefore be simultaneously diagonalized with eigenstates $\ket{k}$, i.e.
\begin{align}
P_\mu \ket{k} = k_\mu \ket{k} \,.
\end{align}
Lorentz transformations map $\ket{k}$ to $\ket{k'}$ with $k_\mu$ and $k'_\mu$ in the same Lorentz orbit. The space spanned by a given state $\ket{k}$ carries a representation of the \emph{little group} which is the stabilizer of $k_\mu$ in the Lorentz group $SO(1,D-1)$, i.e.
\begin{align}
G_k = \{ \Lambda \, | \, \Lambda_\mu{}^\sigma k_\sigma = k_\mu \; \text{and} \; \Lambda \in SO(1,D-1) \} \,.
\end{align}  
The corresponding Lie algebra is called \emph{little algebra}. Unitary irreducible representation of the little algebra uniquely induce unitary irreducible representations of the Poincar\'e algebra. 

In the following, we will illustrate this procedure both for the massive and massless case.

\subsubsection{Massive Case}
Let us consider the case of a Lorentz orbit with $k^2=-m^2$. Following Wigner's method, we consider a particular element of this Lorentz orbit which, for convenience, we choose to be
\begin{align}
k_\mu = (m,0,0,\dots,0) \,.
\end{align}
The corresponding little algebra is given by $\mathfrak{so}(D-1)$. A unitary irreducible representations of this algebra can be encoded by a Young diagram

 \begin{equation}
 \begin{aligned}
&\begin{tabular}{|c|c|c|c|c|}\hline
   $\phantom{.}$&$\phantom{.}$&\multicolumn{2}{|c|}{$~~~~\cdots~~~~\cdots~~~~\cdots~~~~$}&\phantom{.}\\\hline
\end{tabular}~\,m_1\\[-4pt]
&\begin{tabular}{|c|c|c|c|c|}
   $\phantom{.}$&$\phantom{.}$&\multicolumn{2}{|c|}{$~~~~\cdots~~~~$}&$\phantom{.}$\\\hline
\end{tabular}~\,m_2\\[-4.3pt]
&\begin{tabular}{|c|}
   $~~\vdots~~\vdots~~\vdots~~\vdots~~\vdots~~\vdots~~\vdots~~$\\\hline
\end{tabular}\\[-4pt]
&\begin{tabular}{|c|c|c|}
   $\phantom{.}$&$\cdots$&$\phantom{.}$\\\hline
\end{tabular}~\,m_n\;,
 \end{aligned}
  \end{equation}

\noindent which has $n$ rows and there are $m_i$ boxes in the $i$-th row with $m_1 \ge m_2 \ge m_3 \ge \dots \ge m_n$. The $\mathfrak{so}(d-1)$-tensors 
\begin{align}
\phi_{a_{1 1} \dots a_{1 m_1} \,, \,a_{2 1} \cdots a_{2 m_2} \,,\,\dots\,,\,a_{n 1}\dots a_{n m_n}}  \,, && a_{i j} \in \{1,\dots,D-1\} \,,
\end{align} 
form a basis\footnote{This basis is usually referred to as the symmetric basis. One also often considers an antisymmetric basis but we will not do so in the following. See, for example, Appendix~E of \cite{Didenko:2014dwa} for more details.} for the vector space of the representation associated with the Young diagram if 
\begin{itemize}
\item they are completely symmetric in the indices of type $a_{1 i}$, in indices of type $a_{2j}$ and so on.
\item symmetrization of all indices associated with row $i$ with any index associated with row $j>i$ vanishes\footnote{It can be shown that this condition for $j=i+1$ implies all that all symmetrization for $j>i+1$ vanish.}, for example
\begin{align}
\phi_{(a_{1 1} \dots a_{1 m_1} \,, \,a_{2 1}) \dots a_{2 m_2} \,,\,\dots\,,\,a_{n 1}\dots a_{n m_n}} = 0 \,.
\end{align}
\item all traces vanish\footnote{It can be shown that tracelessness in the first class of indices, e.g. $a_{1j}$ with $j \in \{1,\dots,m_1\}$, implies that all other traces vanish. For a proof of this statement, we refer to Appendix E of \cite{Didenko:2014dwa}.}, for example 
\begin{align}
\delta^{a_{11} a_{12}} \, \phi_{a_{1 1} \dots a_{1 m_1} \,, \, a_{2 1} \dots a_{2 m_2} \,,\,\dots\,,\,a_{n 1}\dots a_{n m_n}} = 0 \,.
\end{align}
\end{itemize}
Let us consider the Young diagram with $m_i=0$ for $i=1\dots n$ which is usually denoted by $\bullet$. This corresponds to the trivial representation and the associated vector space is spanned by $\mathfrak{so}(d-1)$-scalars $\phi$. Similarly, the Young diagram $\Box$ corresponds to the vector representation whose representation space is spanned by $\mathfrak{so}(d-1)$-vectors $\phi_a$. More generally, the representation space associated with Young diagrams with $m_1=s$ and $m_i=0$ for $i>1$, 
\begin{equation}
 \begin{aligned}
&\begin{tabular}{|c|c|c|}\hline
   $\phantom{.}$&\multicolumn{1}{|c|}{$~~~~\cdots~~~~$}&\phantom{.}\\\hline
\end{tabular}~\,s \,, \nonumber
\end{aligned}
\end{equation}

 \noindent is spanned by completely symmetric and traceless rank-$s$ tensors $\phi_{a_1 \dots a_s}$. These representations are related by the Wigner method to massive spin-$s$ representations of the Poincar\'e algebra and therefore to massive spin-$s$ particles.

Tensors related to Young diagrams with more than one row are generically neither fully symmetric nor antisymmetric. They are associated with massive mixed symmetry particles which play an important role in string theory. We will however restrict ourselves to the completely symmetric case in the following. Furthermore, we will focus on massless particles only which are discussed in the next section.

\subsubsection{Massless Case}
For the case $k^2=0$, we choose 
\begin{equation}
\label{eq:reprVectorMassless}
k_\mu = (E,0,\dots,0,E) \,.
\end{equation}
Lorentz transformations generated by the $\mathfrak{so}(D-2)$ subalgebra leave this vector invariant. All massless particles in nature transform in a representation of the Poincar\'e algebra induced by this algebra.\footnote{There is a subtlety here: the maximal subalgebra of the Lorentz algebra leaving \eqref{eq:reprVectorMassless} invariant is actually $\mathfrak{iso}(D-2)$. However, in nature, we exclusively observe particles which transform in representations for which only the generators of the subalgebra $\mathfrak{so}(D-2) \subset \mathfrak{iso}(D-2)$ are non-trivially realized.\\Representations which realize the full $\mathfrak{iso}(D-2)$ algebra non-trivially could again by studied using the Wigner method. They would lead to a momentum-like continuous quantum number which is usually referred to as 'continuous spin'. Such a quantum number is not observed in nature and we will therefore not consider these representations in the following.} Therefore, the representation theory for the massless case is the same as for massive representations in one dimension lower. In particular, there is a spin-$s$ representation whose vector space is spanned by completely symmetric and traceless $\mathfrak{so}(D-2)$-tensors $\phi_{b_1 \dots b_s}$ with $b_i \in \{ 1 \dots D-2 \}$. These representations are related to massless spin-$s$ particles.

\subsubsection{Counting Degrees of Freedom}
\label{sec:starsandbars}
The degrees of freedom of a massless spin-$s$ particle is given by the dimension of the massless spin-$s$ representation discussed in the last section. As explained, its representation space is spanned by $\mathfrak{so}(D-2)$-tensors 
\begin{align}
\phi_{b_1 \dots b_s} \,,
\end{align}
which are completely symmetric and traceless. So counting the degrees of freedom of a massless spin-$s$ particle reduces to counting the independent components of completely symmetric and traceless $\mathfrak{so}(D-2)$-tensors.

Let us first neglect the trace constraint and consider fully symmetric $\mathfrak{so}(D-2)$-tensors. Their independent components can be conveniently determined using the \emph{stars and bars trick}. For this, first notice that because all indices $b_i$ are symmetric their order is of no importance. We will therefore represent each index by a star $\star$. The different values that these indices can take are represented by $D-2$ 'buckets' separated by $D-3$ bars $|$. As an example, for $D=5$ and $s=3$  we represent $\phi_{111}$ by
\begin{equation}
\star \star \star| \, |
\end{equation}
and $\phi_{123}=\phi_{213}=\dots$ by
\begin{equation}
\star  | \star | \star \,.
\end{equation}
The total number of possible combinations of $D-3$ bars and $s$ stars is given by
\begin{align}
\binom{D-3+s}{s}
\end{align}
and is therefore equal to the independent components of a fully symmetric rank-$s$ $\mathfrak{so}(D-2)$-tensor. To take into account the trace constraint, one has to subtract its independent degrees of freedom. Because taking the trace of a tensor reduces the number of free indices by two, we conclude that a completely symmetric and traceless $\mathfrak{so}(D-2)$-tensor of rank-$s$ has 
\begin{align}
\label{eq:correctDOF}
\binom{D-3+s}{s} - \binom{D-5+s}{s-2}
\end{align}
independent components which is precisely \eqref{eq:numOfDegOfFreed}.

%----------------------------------------------------------------
\section{Propagating Degrees of Freedom of Fronsdal Field}
\label{sec:DOFofFronsdal}
In the last section, it was discussed that the number of degrees of freedom of a massless spin-$s$ field is \eqref{eq:numOfDegOfFreed}. In this section, we will try to show that the Fronsdal field $\phi_{\mu_1 \dots \mu_s}$ indeed carries these degrees of freedom. 

As we will prove in the following, the Fronsdal equation describes the propagation of a massless spin-$s$ particle if we require double-tracelessness for the Fronsdal field
\begin{align}
\label{eq:doubleTracelessness}
\phi_{\mu_1 \dots \mu_{s-4}}{}^{\sigma}{}_{\sigma}{}^{\kappa}{}_{\kappa} = 0 \,.
\end{align}
Note that the double-trace constraint \eqref{eq:doubleTracelessness} is gauge invariant as its gauge variation necessarily involves a trace of the gauge parameter which vanishes, see \eqref{eq:tracelessnessOfGaugeParameter}.

In order to show that the double-traceless Fronsdal field indeed propagates the correct degrees of freedom, it is convenient to choose \emph{de~Donder gauge}
\begin{align}
\label{eq:deDonderGaugeCondition}
D_{\mu_1 \dots \mu_{s-1}}=\partial^\sigma \phi_{\mu_1 \dots \mu_{s-1} \sigma} - \frac12 \, \partial_{(\mu_1} \phi_{\mu_2 \dots \mu_{s-1}) \sigma}{}^\sigma = 0 \,.
\end{align}
For later purposes, we note that the de~Donder tensor $D_{\mu_1 \dots \mu_{s-1}}$ is traceless as you are invited to check in the following problem. 

\begin{problem} Show that the de~Donder tensor $D_{\mu_1 \dots \mu_{s-1}}$  is traceless.
\end{problem}

\noindent By calculating the gauge variation of the de~Donder tensor, 
\begin{align}
\delta D_{\mu_1 \dots \mu_{s-1}} = \Box \xi_{\mu_1 \dots \mu_{s-1}} \,,
\end{align}
we see that \eqref{eq:deDonderGaugeCondition} does not fix the gauge completely. The residual gauge freedom is given by gauge parameters which obey $\Box \xi_{\mu_1 \dots \mu_{s-1}} = 0$.

\begin{problem} Calculate the gauge variation of the de~Donder tensor $D_{\mu_1 \dots \mu_{s-1}}$.
\end{problem}

\noindent In de~Donder gauge, the Fronsdal equation \eqref{eq:fronsdalEquation} becomes a simple wave equation
\begin{equation}
\Box \phi_{\mu_1 \dots \mu_s}  = 0 \,,
\end{equation}
as can be easily shown by observing that $D_{\mu_1 \dots \mu_{s-1}}=0$ also implies that the following tensor vanishes\footnote{There is no factor of $\frac12$ in the last term as for the symmetrization on the left hand side of \eqref{eq:derivDeDonder} one needs $s$ permutations. The last term in the de~Donder tensor contains $s-1$ permutations. On the other hand, the last summand on the right hand side of \eqref{eq:derivDeDonder} consists of $\frac{s(s-1)}{2}$ permutations. The factor $\frac12$ in \eqref{eq:deDonderGaugeCondition} balances this mismatch.}
\begin{align}
\label{eq:derivDeDonder}
\partial_{(\mu_1} D_{\mu_2 \dots \mu_{s})} = \partial_{(\mu_1} \partial^\sigma \phi_{\mu_2 \dots \mu_s) \sigma} + \partial_{(\mu_1} \partial_{\mu_2} \phi_{\mu_3 \dots \mu_s)\sigma}{}^{\sigma}  \,. 
\end{align}
But this tensor precisely coincides with the last two terms in the Fronsdal equation \eqref{eq:fronsdalEquation}.
In this gauge, the solution for the Fronsdal equation therefore takes the form
\begin{align}
\phi_{\mu_1 \dots \mu_s}(x) = \int \text{d}^D k \; e^{i kx} \, e_{\mu_1 \dots \mu_s}(k) 
\end{align}
with $k^2 = 0$. The completely symmetric tensor $e_{\mu_1 \dots \mu_s}$ is double-traceless and therefore has
\begin{align}
\label{eq:indepCompOfFronsdal}
\binom{D-1+s}{s} - \binom{D-5+s}{s-4}
\end{align}
independent components - as can be verified by the stars and bars method along similar lines as in Section~\ref{sec:starsandbars}. But some of these components are related by the gauge condition \eqref{eq:deDonderGaugeCondition} which, since the de Donder tensor is traceless, imposes 
\begin{align}
\binom{D-2+s}{s-1} - \binom{D-4+s}{s-3}
\end{align}
conditions on $e_{\mu_1 \dots \mu_s}$. Furthermore, the residual gauge symmetry $\Box \xi_{\mu_1 \dots \mu_{s-1}}=0$ is solved by
\begin{align}
\xi_{\mu_1 \dots \mu_{s-1}}(x) = \int \text{d}^D x \, e^{i kx} \, \tilde{\xi}_{\mu_1 \dots \mu_{s-1}}(k) \,
\end{align}
with $k^2 = 0$. Since $\tilde{\xi}$ is traceless, this allows us to eliminate 
\begin{align}
\binom{D-2+s}{s-1} - \binom{D-4+s}{s-3}
\end{align}
components from $e_{\mu_1 \dots \mu_s}$. So in total, we are left with 
\begin{align}
\binom{D-1+s}{s} - \binom{D-5+s}{s-4} - 2 \left\{ \binom{D-2+s}{s-1} - \binom{D-4+s}{s-3}\right\}
\end{align}
degrees of freedom. One can easily check that this number precisely agrees with \eqref{eq:numOfDegOfFreed} and therefore the Fronsdal equation indeed propagates the correct number of degrees of freedom.

Note that for this proof, it was essential that the double-trace constraint was imposed.\footnote{It is important to emphasize that we have only proven that the double-tracelessness of the Fronsdal field is a sufficient condition for the propagation of the correct degrees of freedom. As is discussed in Appendix~\ref{app:doubleTrace}, it is \emph{not} a necessary condition. If the double-trace constraint is not imposed, the second and higher-traces of the Fronsdal field vanish on-shell (provided that we impose suitable boundary conditions) and therefore do not propagate any additional degrees of freedom . In this case however, the Fronsdal equation cannot be derived from a gauge-invariant action. For this reason, one usually imposes the double-trace constraint.} The Fronsdal equation therefore indeed describes the propagation of a massless spin-$s$ field.

%----------------------------------------------------------------
\section{Action}
\label{sec:action}

Now that we have shown that the Fronsdal equation \eqref{eq:fronsdalEquation} indeed describes the propagation of a massless spin-$s$ particle, it is natural to construct the corresponding Fronsdal action whose equations of motion are equivalent to the Fronsdal equation \eqref{eq:fronsdalEquation}. As we will show in the following, this action is given by
\begin{align}
\label{eq:fronsdalAction}
S = \frac12 \int \text{d}^D x \;
\phi^{\mu_1 \dots \mu_s} \, \mathcal{F}_{\mu_1 \dots \mu_s} \,,
\end{align}
where we have defined
\begin{align}
\label{eq:definitionCurlyF}
\mathcal{F}_{\mu_1 \dots \mu_s} =  F_{\mu_1 \dots \mu_s} - \frac12 \, \eta_{( \mu_1 \mu_2} F_{\mu_3 \dots \mu_s)\sigma}{}^\sigma \,,
\end{align}
with $F_{\mu_1 \dots \mu_s}$ denoting the Fronsdal tensor defined in \eqref{eq:fronsdalEquation}. As you are invited to check in Problem 4, the Fronsdal action \eqref{eq:fronsdalAction} is symmetric in the sense that 
\begin{align}
\int \text{d}^D x \;
\phi^{\mu_1 \dots \mu_s} \, \mathcal{F}_{\mu_1 \dots \mu_s}(\psi) = \int \text{d}^D x \;
\psi^{\mu_1 \dots \mu_s} \, \mathcal{F}_{\mu_1 \dots \mu_s}(\phi) \,, \label{eq:symmetryPropertyAction}
\end{align}
where we have to impose suitable boundary conditions such that all total derivatives in the integrand lead to vanishing contributions to the action. Using this symmetry, one can easily vary the action 
\begin{align}
\delta S &= \frac12 \int \text{d}^D x \; \big\{
\delta \phi^{\mu_1 \dots \mu_s} \, \mathcal{F}_{\mu_1 \dots \mu_s}(\phi) + \phi^{\mu_1 \dots \mu_s} \, \mathcal{F}_{\mu_1 \dots \mu_s}(\delta \phi) \big\}  \nonumber \\
&= \hphantom{\frac12}\int \text{d}^D x \; 
\delta \phi^{\mu_1 \dots \mu_s} \, \mathcal{F}_{\mu_1 \dots \mu_s}(\phi) \label{eq:variationOfFronsdalAction}
\end{align}
and obtain the corresponding equation of motion
\begin{equation}
\label{eq:eomAction}
\mathcal{F}_{\mu_1 \dots \mu_s} = 0 \,.
\end{equation}
Note that this is not the Fronsdal equation \eqref{eq:fronsdalEquation}. However, one can easily show that \eqref{eq:eomAction} is equivalent to the Fronsdal equation \eqref{eq:fronsdalEquation} by taking the trace of \eqref{eq:definitionCurlyF} which gives 
\begin{align}
\eta^{\mu_{s-1}\mu_s} \mathcal{F}_{\mu_1 \dots \mu_s} \propto F_{\mu_1 \dots \mu_{s-2} \sigma}{}^\sigma \,. \label{eq:vanishingTraceByEoM}
\end{align}
This result will be checked in Problem 5. Thus the equation of motion \eqref{eq:eomAction} implies that the trace of the Fronsdal tensor vanishes, i.e. $F_{\mu_1 \dots \mu_{s-2} \sigma}{}^\sigma = 0$. The Fronsdal tensor $F$ and the tensor $\mathcal{F}$ only differ by a trace term - as can be seen by comparing with the definition of $\mathcal{F}$ in \eqref{eq:definitionCurlyF}. We therefore conclude that the equation of motion \eqref{eq:eomAction} indeed implies the Fronsdal equation \eqref{eq:fronsdalEquation}.
\begin{problem}
Check that the symmetry property \eqref{eq:symmetryPropertyAction} indeed holds. \emph{Hint: First consider the case $s=3$.}
\end{problem}

\begin{problem}
Prove the relation \eqref{eq:vanishingTraceByEoM}.
\end{problem}

Along similar lines, one can also show that the Fronsdal action is gauge invariant under $\delta \phi_{\mu_1 \dots \mu_s}= \partial_{(\mu_1} \xi_{\mu_2 \dots \mu_s)}$ with traceless gauge parameter $\xi$. The variation \eqref{eq:variationOfFronsdalAction} and partial integration implies that
\begin{align}
\delta S \propto \int \text{d}^D x \; 
 \xi^{\mu_2 \dots \mu_s} \, \partial^{\mu_1} \mathcal{F}_{\mu_1 \dots \mu_s}
\end{align}
Using the definition of $\mathcal{F}$ of \eqref{eq:definitionCurlyF}, the integrand is given by
\begin{align}
\xi^{\mu_1 \dots \mu_{s-1}} \left( \partial^\sigma F_{\mu_1 \dots \mu_{s-1} \sigma} - \frac12 \partial^\sigma \eta_{(\sigma \mu_1} F_{\mu_2 \dots \mu_{s-1})\lambda}{}^\lambda\right) \,, \label{eq:integrandOfGaugeVariationOfFronsdal}
\end{align}
where we have suitably relabeled the indices.
Since the gauge parameter is traceless, the last term only contributes if the metric $\eta$ carries the index $\sigma$. Therefore, the integrand is given by
\begin{align}
\xi^{\mu_1 \dots \mu_{s-1}} \left( \partial^\sigma F_{\mu_1 \dots \mu_{s-1} \sigma} - \frac12 \partial_{(\mu_1} F_{\mu_2 \dots \mu_{s-1})\lambda}{}^\lambda\right) \,. 
\end{align}
By an explicit calculation, one can then show that the expression in the bracket vanishes
\begin{align}
\partial^\sigma F_{\mu_1 \dots \mu_{s-1} \sigma} - \frac12 \partial_{(\mu_1} F_{\mu_2 \dots \mu_{s-1})\lambda}{}^\lambda \equiv 0 \label{eq:binachiIdent}
\end{align}
and therefore the action is gauge invariant. Equation \eqref{eq:binachiIdent} is called the \emph{Bianchi identity} and will be proven in the following exercise.
\begin{problem}
Show that \eqref{eq:binachiIdent} indeed holds.
\end{problem}

In summary, it was shown in this section that there exists a gauge invariant action \eqref{eq:fronsdalAction} whose equation of motion are equivalent to the Fronsdal equation \eqref{eq:fronsdalEquation}. Although we will not prove this statement, it is important to note that the Fronsdal action is unique - up to partial integration and an overall constant.

%----------------------------------------------------------------
\section{AdS backgrounds}
\label{sec:AdSbackgrounds}
So far, we have considered higher-spin fields propagating on a Minkowski background. As we will now discuss, one can also consistently define a Fronsdal equation for dS and AdS background geometries. For concreteness, we will only explain the latter case in detail.

Naively, one could hope to obtain the Fronsdal equation for AdS from the one for flat space  \eqref{eq:fronsdalEquation} by replacing partial derivatives $\partial$ by the covariant derivatives $\nabla$ of AdS space. One then needs to check whether the resulting equation is gauge invariant under 
\begin{equation}
\label{eq:gaugeTrafoAds}
\delta \phi_{\mu_1 \dots \mu_s} = \nabla_{(\mu_1} \xi_{\mu_2 \dots \mu_s)} \; \; \; \textrm{with} \; \;  g^{\sigma \kappa} \xi_{\sigma \kappa \mu_1 \dots \mu_{s-3}}=0 \,,
\end{equation}
where $g_{\mu \nu}$ denotes the metric of AdS space.
The calculation would follow similar lines as for the flat case in Problem 1 with the additional complication that the covariant derivatives no longer commute, e.g.
\begin{equation}
\label{eq:commCovDeriv}
[\nabla_\mu, \nabla_\nu] v_\rho = -\frac{1}{l^2} \, ( g_{\mu \rho} \, v_\nu - g_{\nu \rho} \, v_\mu) \,,
\end{equation}
where $l$ is the AdS radius. This in turn leads to additional terms in the gauge variation of the action that do not cancel out. Luckily, there is an easy way to fix this: one just adds two additional terms to the action which precisely cancel the contributions of the commutators. The resulting Fronsdal equation is then given by\footnote{The factor of $\frac12$ in the last term of the first line is due to our symmetrization convention as explained in Appendix A.}
\begin{align}
\label{eq:fronsdalEquationAdS}
\Box \phi_{\mu_1 \dots \mu_s} - \nabla_{(\mu_1} \nabla^\sigma &\phi_{\mu_2 \dots \mu_s) \sigma} + \frac12 \, \nabla_{(\mu_1} \nabla_{\mu_2} \phi_{\mu_3 \dots \mu_s)\sigma}{}^{\sigma} \nonumber \\ 
&- \frac{1}{l^2} \, m_s^2 \, \phi_{\mu_1 \dots \mu_s} - \frac{2}{l^2} \, g_{(\mu_1 \mu_2} \phi_{\mu_3 \dots \mu_s)\sigma}{}^\sigma = 0 \,,
\end{align}
where $m_s^2 = s^2+s(D-6)-2(D-3)$. Note that the mass-like term proportional to $m_s^2$ in the equation of motion is required for gauge invariance. This is different to the Minkowski case where a mass term would break gauge invariance.\footnote{The concept of mass is a bit subtle in AdS space as $P^2$ is not a quadratic Casimir of the AdS isometry algebra, where $P_\mu$ is the generalized translation operator of the AdS isometry algebra.} 
\begin{problem}
Show that \eqref{eq:fronsdalEquationAdS} is invariant under \eqref{eq:gaugeTrafoAds}. Warning: This problem is more difficult. If you find it hard to solve read the solutions and make sure that you can follow the calculation.
\end{problem}

The Fronsdal equation for dS space can be obtained by flipping the sign of the cosmological constant $\Lambda \sim \frac{1}{l^2}$. One can also find a suitable generalization of the Fronsdal action \eqref{eq:fronsdalAction} for (A)dS backgrounds. We refer to Section 2 of \cite{Didenko:2014dwa} for a discussion of this.

Our proof for the gauge invariance of \eqref{eq:fronsdalEquationAdS} heavily relies on the fact that the commutators of covariant derivatives take the form \eqref{eq:commCovDeriv} and therefore only holds for maximally symmetric spacetimes. For generic backgrounds, the gauge variation of the first line in \eqref{eq:fronsdalEquationAdS} is schematically of the form
\begin{equation}
R_{\dots} (\nabla \xi){}^{\dots} + (\nabla R){}_{\dots} \xi^{\dots} \,,
\end{equation}
where $\xi$ and $R$ are the spin-$s$ gauge parameter and the background Riemann tensor respectively and the ellipsis schematically denote various contractions of indices. The last term can not be canceled by adding additional terms to the Fronsdal equation because the gauge parameter arises without a covariant derivative acting on it. Therefore for generic backgrounds, one can not construct a gauge invariant (generalization of the) Fronsdal equation.

From our discussion in this section, it follows that maximal symmetry of a spacetime background is a sufficient condition for the existence of a gauge invariant Fronsdal equation. However, to the best of my knowledge, a necessary and sufficient condition is not yet known.

%----------------------------------------------------------------
\section{Outlook}
In these lectures, we have discussed the Fronsdal equation in detail. Gauge invariance of the Fronsdal equation imposes tracelessness of its corresponding gauge parameter, as was discussed in Section~\ref{sec:FronsdalEquation}. We then showed in Section~\ref{sec:DOFofFronsdal} that the Fronsdal equation indeed describes the correct degrees of freedom. For our proof, it was essential to impose that the Fronsdal field is double-traceless. In Section~\ref{sec:action}, we then presented an action whose Euler-Lagrange equations are equivalent to the Fronsdal equation. Until this point, our discussion was valid only for flat backgrounds. In Section~\ref{sec:AdSbackgrounds}, we then generalized the Fronsdal equation to the other maximally symmetric backgrounds, i.e. AdS and dS.

For $s=2$, the Fronsdal equation reduces to the linearized Einstein equations. It is natural to ask if there is also a generalization of the full Einstein equations for higher-spin fields or, put differently, if there exist non-linear field equations which reduce to the Fronsdal equations upon linearization around a given background. For AdS and dS backgrounds, such equations were indeed found by Vasiliev and collaborators \cite{Prokushkin:1998bq,Vasiliev:1999ba}. For flat backgrounds, it is widely believed that no such equations exist.

Vasiliev equations are formulated in a highly non-standard manner using an infinite number of auxiliary fields and coordinates. A detailed understanding of its physical implications is still an active area of current research.

Both four- and three-dimensional Vasiliev theory are of particular interest as they arise as bulk duals of particularly simple conformal field theories. A certain type of four-dimensional Vasiliev theory is dual to the free $O(N)$ vector-model, i.e. $N$ free bosons which transform in the fundamental representation of the global $O(N)$ symmetry \cite{Klebanov:2002ja,Sezgin:2003pt}. Three-dimensional Vasiliev theories are dual to a certain generalization of two-dimensional minimal models \cite{Gaberdiel:2012uj}. These dualities have generated considerable attention over the last years and provide an interesting class of AdS/CFT dualities from which one might hope to understand the underlying mechanisms of these correspondences better. See, for example, \cite{Giombi:2016ejx} for a review of the four-dimensional and \cite{Gaberdiel:2012uj} for the three-dimensional case.

String theory contains an infinite tower of massive higher-spin fields with masses $M^2 \sim l_s^{-2}$, where $l_s$ is the string length. One considers typically the point particle limit for which $l_s$ is taken to be small compared to the length scale we are interested in. In this regime, the higher-spin fields become very massive and are therefore irrelevant for low-energy physics. However, there is also the opposite limit of $l_s$ much greater than the physical length scale. In this tensionless limit, all higher-spin fields are massless and the theory therefore possesses a huge higher-spin gauge symmetry. It is widely believed that this is the underlying gauge algebra of string theory and by higgsing this gauge symmetry the infinite tower of higher-spin fields becomes massive. Over the last years, this Higgs mechanism has become a very active and exciting field of research \cite{Chang:2012kt,Gaberdiel:2014cha,Gaberdiel:2015wpo}. This was achieved by comparing the dual conformal field theories of particular higher-spin and string theories (in the tensionless limit and on certain backgrounds).

Given all these exciting applications, higher-spin theories are a topic worth studying and hopefully these lectures will help the reader in learning more about the subject.

%----------------------------------------------------------------
\section*{Acknowlegements}
I want to thank Andrea Campoleoni, Stefan Fredenhagen, Alexander Kegeles, Gustavo Lucena Gomez, Evgeny Skvortsov, Charlotte Sleight,  Rakibur Rahman and Karapet Mkrtchyan for useful discussions. In particular, I am very much indebted to Rakib and Karapet for patiently answering various questions that I had. I want to thank Christian Northe for carefully reading my draft and pointing out various mistakes. Last but not least, I want to thank the organizers of the Modave school for inviting me and the participants for making the week such a wonderful experience.

%----------------------------------------------------------------
\appendix
\section{Conventions}
\label{app:conventions}
Throughout these lectures, we use symmetrization conventions which involve all necessary permutations without any additional factors, for example
\begin{align}
\partial_{(\mu} \xi_{\nu)} = \partial_\mu \xi_\nu + \partial_\nu \xi_\mu \,.
\end{align}
Similarly, for the fully symmetric tensor $\xi^{\mu_1 \dots \mu_{s-1}}$, we have
\begin{align}
\partial^{(\mu_1}\xi^{\mu_2 \dots \mu_{s})} = \partial^{\mu_1} \xi^{\mu_2 \dots \mu_s} + \partial^{\mu_2} \xi^{\mu_1 \mu_3 \mu_4 \dots \mu_s} + \dots + \partial^{\mu_s} \xi^{\mu_1 \dots \mu_{s-1}} \,,
\end{align}
so in total $s$ permutations. As a last example, we note that $\partial_{(\mu_1} \partial_{\mu_2} \phi'_{\mu_3 \dots \mu_{s-2})}$ involves $\binom{s}{2}$ terms, e.g. for $s=3$
\begin{align}
\partial_{(\mu_1} \partial_{\mu_2} \phi'_{\mu_3 \dots \mu_s)} = \partial_{\mu_1} \partial_{\mu_2} \phi'_{\mu_3} + \partial_{\mu_1} \partial_{\mu_3} \phi'_{\mu_2} + \partial_{\mu_2} \partial_{\mu_3} \phi'_{\mu_1} \,,
\end{align}
whereas $\nabla_{(\mu_1} \nabla_{\mu_2} \phi'_{\mu_3 \dots \mu_{s})}$ involves $s(s-1)$ terms because the covariant derivatives do not commute, e.g. for $s=3$
\begin{align}
\nabla_{(\mu_1} \nabla_{\mu_2} \phi'_{\mu_3 \dots \mu_3)} = &\;\hphantom{+} \nabla_{\mu_1} \nabla_{\mu_2} \phi'_{\mu_3} + \nabla_{\mu_2} \nabla_{\mu_1} \phi'_{\mu_3} \\  &+ \nabla_{\mu_1} \nabla_{\mu_3} \phi'_{\mu_2} + \nabla_{\mu_3} \nabla_{\mu_1} \phi'_{\mu_2} \\& + \nabla_{\mu_2} \nabla_{\mu_3} \phi'_{\mu_1} + \nabla_{\mu_3} \nabla_{\mu_2} \phi'_{\mu_1} \,.
\end{align}
This mismatch in the number of permutations also explains the relative factor of $\frac12$ in the last term of the first line in \eqref{eq:fronsdalEquationAdS} with respect to \eqref{eq:fronsdalEquation}.

 Although these conventions might look rather cumbersome on first sight, they are convenient as they tend to lead to a lower number of explicit factors in the equations.  

 By $\phi'_{\mu_1 \dots \mu_{s-2}}$ we denote the trace of the Fronsdal field. We use similar notation for other tensors as well.

%----------------------------------------------------------------
\section{More on the Double-trace Constraint}
\label{app:doubleTrace}
In Section~\ref{sec:DOFofFronsdal}, we have seen that the Fronsdal field propagates the correct degrees of freedom \eqref{eq:correctDOF} of a massless spin-$s$ field. Our proof relied on the fact that the Fronsdal field is double-traceless \eqref{eq:doubleTracelessness}. In this appendix, we will show that the Fronsdal field propagates the correct degrees of freedom \eqref{eq:correctDOF} even without imposing the double-trace constraint (provided that we impose suitable boundary conditions). This point is often stated incorrectly in the literature.\footnote{I want to thank Karapet Mkrtchyan for patiently explaining this subtlety to me. To the best of my knowledge, this observation was first made in \cite{Francia:2010ap} - see footnote 2 on page 2 of this reference.} 

In order to see this, let us assume that the double-trace constraint \eqref{eq:doubleTracelessness} is \emph{not} imposed on the Fronsdal field. As a result, the Fronsdal equation \eqref{eq:fronsdalEquation} contains additional components due to non-vanishing higher traces\footnote{By higher traces of a fully symmetric tensor $t_{\mu_1 \dots \mu_s}$, we mean 
\begin{align*}
t_{\mu_1 \dots \mu_{s-4}}{}^{\lambda \sigma}{}_{\lambda \sigma} \,, && t_{\mu_1 \dots \mu_{s-6}}{}^{\kappa \lambda \sigma}{}_{\kappa \lambda \sigma}\,, && \dots \,.
\end{align*}
}. Furthermore the Bianchi identity \eqref{eq:binachiIdent} is modified to
\begin{align}
\partial^\sigma F_{\mu_1 \dots \mu_{s-1} \sigma} - \frac12 \partial_{(\mu_1} F_{\mu_2 \dots \mu_{s-1})\lambda}{}^\lambda  = - \frac32 \, \partial_{(\mu_1} \partial_{\mu_2} \partial_{\mu_3} \phi_{\mu_4 \dots \mu_{s-1})}{}^{\lambda \sigma}{}_{\lambda \sigma} \,. \label{eq:modifiedBinachiIdent}
\end{align}
\begin{problem}
Show that the modified Bianchi identity \eqref{eq:modifiedBinachiIdent} indeed holds.
\end{problem}
We therefore conclude that imposing the Fronsdal equation $F_{\mu_1 \dots \mu_s}=0$ implies that 
\begin{align}
\partial_{(\mu_1} \partial_{\mu_2} \partial_{\mu_3} \phi_{\mu_4 \dots \mu_{s-1})}{}^{\lambda \sigma}{}_{\lambda \sigma} = 0 \,.
\end{align}
But this differential equation has only polynomial solutions. For example for $s=4$, the most general solution is given by $\phi^{\lambda \sigma}{}_{\lambda \sigma} = c + c_{\mu} x^\mu + c_{\mu \nu} x^\mu x^\nu$. If we require that the Fronsdal field vanishes at infinity, the only solution is given by $\phi_{\mu_1 \dots \mu_{s-4}}{}^{\lambda \sigma}{}_{\lambda \sigma} = 0$. Therefore, the Fronsdal field is now double-traceless on-shell and, as a result, its higher traces do not carry any degrees of freedom. But for the other components of the Fronsdal field, we can repeat exactly the same argument as in Section~\ref{sec:DOFofFronsdal} to show that they carry the expected number of degrees of freedom \eqref{eq:numOfDegOfFreed} of a massless spin-$s$ field. 

We also immediately conclude that the Fronsdal action\footnote{By Fronsdal action, we mean \eqref{eq:fronsdalAction} where we use the definition of $\mathcal{F}$ given in \eqref{eq:definitionCurlyF} with the Fronsdal operator defined in \eqref{eq:fronsdalEquation}. The latter is now no longer double-traceless as we have not imposed a double-tracelessness constraint on the Fronsdal field.} \eqref{eq:fronsdalAction} is no longer gauge invariant as this required the Bianchi identity \eqref{eq:binachiIdent} which is now modified. One could easily restore gauge invariance by modifying the definition \eqref{eq:definitionCurlyF} of $\mathcal{F}$ to
\begin{equation}
\mathcal{F}_{\mu_1 \dots \mu_s} \to \mathcal{F}_{\mu_1 \dots \mu_s} + \frac12 \, \eta_{\sigma (\mu_1} \partial_{\mu_2} \partial_{\mu_3} \phi''_{\mu_4 \dots \mu_{s})} \,.
\end{equation}
The additional term precisely cancels the term on the right hand side of the modified Bianchi identity \eqref{eq:modifiedBinachiIdent}. However, the resulting equations of motion are no longer equivalent to the Fronsdal equations and propagate ghost degrees of freedom in addition to a massless spin-$s$ field.

In summary, we have seen in this appendix that the Fronsdal equation propagates the correct degrees of freedom \eqref{eq:correctDOF} even without imposing double-tracelessness of the Fronsdal field but cannot be derived from a gauge invariant action in this case. Double-tracelessness of the Fronsdal field is therefore a necessary condition for gauge invariance of the Fronsdal action but not for the propagation of the correct degrees of freedom of the Fronsdal equation (for which it is however a sufficient condition as we have seen in Section~\ref{sec:DOFofFronsdal}).

Let us also mention in passing that there are also formulations of actions and equations of motion which do not require a trace constraint on the gauge parameter at the price of introducing non-localities or auxiliary fields (see \cite{Francia:2005bu} and references therein).

%----------------------------------------------------------------
\section{Solutions}

%-----------------------------------------
\subsection*{Problem 1}
\noindent There are various ways to check this but one of the most simplest is using the following trick\footnote{I want to thank Rakibur Rahman for pointing this out.}: one introduces auxiliary constant vectors $u^\mu$. Using these, we can define a generating function for the Fronsdal field
\begin{align}
\label{eq:fronsdalGenFunc}
\phi(u) = \frac{1}{s!} \, \phi_{\mu_1 \dots \mu_s} \, u^{\mu_1} \dots u^{\mu_s} \,.
\end{align}
Furthermore, we will use the notation $d_\mu \equiv \partial^u_\mu$. With these definitions, the Fronsdal equation can be rewritten as
\begin{align}
\label{eq:fronsdalEqGenFunc}
F \phi(u) = (\Box - u\partial \, d \partial + \frac12 \, u\partial \, u \partial \, d^2 ) \phi(u) = 0 \,, 
\end{align}
where we use for example the notation $u\partial = u^\sigma \partial_\sigma$ for contractions and similar for other terms. The equivalence to the Fronsdal equation can be seen by plugging \eqref{eq:fronsdalGenFunc} in \eqref{eq:fronsdalEqGenFunc} and performing all differentiations with respect to $u$. One then obtains
\begin{align*}
&\frac{1}{s!} \left( \Box \phi_{\mu_1 \dots \mu_s} - s \, \partial^\sigma \partial_{\mu_1} \phi_{\mu_2 \dots \mu_s \sigma} + \frac{s (s-1)}{2} \, \partial_{\mu_1} \partial_{\mu_2} \phi'_{\mu_3 \dots \mu_s} \right)  u^{\mu_1} \dots u^{\mu_s} \\
=& \frac{1}{s!} \left( \Box \phi_{\mu_1 \dots \mu_s} - \, \partial^\sigma \partial_{(\mu_1} \phi_{\mu_2 \dots \mu_s) \sigma} + \partial_{(\mu_1} \partial_{\mu_2} \phi'_{\mu_3 \dots \mu_s)} \right)  u^{\mu_1} \dots u^{\mu_s}\,.
\end{align*}
The term is the bracket of the last line is the Fronsdal equation \eqref{eq:fronsdalEquation}. To obtain this line, we have made use of our symmetrization conventions explained in Appendix~A.

Along similar lines, one can check that the gauge variation of the Fronsdal field in this language is given by
\begin{align}
\delta \phi(u) = u\partial \, \xi(u) \,,
\end{align}
where $\xi(u)=\tfrac{1}{(s-1)!} \, \xi_{\mu_1 \dots \mu_{s-1}} \, u^{\mu_1} \dots u^{\mu_{s-1}} $. Therefore, the variation of the Fronsdal equation is given by
\begin{align*}
F \delta \phi(u) = \left(\Box - u\partial \, d \partial + \frac12 \, u\partial \, u \partial \, d^2 \right) u\partial \, \xi(u)
\end{align*}
Using the fact that $[d^2,u\partial]=2 \, d \partial$, this can be rewritten as
\begin{align*}
F \delta \phi =\left(\Box \, u\partial - u \partial \, [d \partial, u \partial] + \frac12 \, (u\partial)^3 \, d^2 \right) \xi(u) \,.
\end{align*}
With $[d\partial, u\partial] = \Box$, we arrive at
\begin{align*}
F \delta \phi(u) &= \left( [\Box,u\partial] + \frac12 (u\partial)^3 \, d^2 \right) \xi(u) \\
&= \frac12 (u\partial)^3 \, d^2 \, \xi(u) \,,
\end{align*}
where we have used that the commutator in the first line vanishes since the spacetime partial derivatives commute.

One can now convert this expression back into index notation as follows
\begin{align*}
\frac1{s!} \, \delta F_{\mu_1 \dots \mu_s} \, u^{\mu_1} \dots u^{\mu_s} &= \frac1{s!} \left( \tfrac12 \, s (s-1) (s-2) \, \partial_{\mu_1} \partial_{\mu_2} \partial_{\mu_3} \xi'_{\mu_4 \dots \mu_s} \right) u^{\mu_1} \dots u^{\mu_s} \\
&= \frac1{s!} \left( 3 \, \partial_{(\mu_1} \partial_{\mu_2} \partial_{\mu_3} \xi'_{\mu_4 \dots \mu_s)} \right) u^{\mu_1} \dots u^{\mu_s}
\end{align*}
We thus conclude that $\delta F_{\mu_1 \dots \mu_s} = 3  \, \partial_{(\mu_1} \partial_{\mu_2} \partial_{\mu_3} \xi'_{\mu_4 \dots \mu_s)}$.

The formalism introduced above has the advantage that it takes automatically care of all symmetrization factors and also generalizes straightforwardly to other backgrounds - as we will see in a later exercise.

%-----------------------------------------
\subsection*{Problem 2}
\noindent By taking the trace and using the fact that the Fronsdal field is double-traceless, one obtains
\begin{align*}
g^{\mu_{s-2} \mu_{s-1}} \, D_{\mu_1 \dots \mu_{s-1}}=&g^{\mu_{s-2} \mu_{s-1}} \left( \partial^\sigma \phi_{\mu_1 \dots \mu_{s-1} \sigma} - \tfrac12 \, \left( \partial_{\mu_1} \phi'_{\mu_2 \dots \mu_{s-1}} + \partial_{\mu_2} \phi'_{\mu_1\mu_3 \dots \mu_{s-1}} + \dots +\partial_{\mu_{s-1}} \phi'_{\mu_1 \dots \mu_{s-2}} \right) \right) \\
=& \partial^\sigma \phi'_{\mu_1 \dots \mu_{s-3} \sigma} - \partial^\sigma \phi'_{\mu_1 \dots \mu_{s-3} \sigma} = 0 \,.
\end{align*}

%-----------------------------------------
\subsection*{Problem 3}
\noindent The gauge variation of the de Donder tensor is given by
\begin{align*}
\delta D_{\mu_1 \dots \mu_{s-1}} =& \partial^\lambda \delta \phi_{\lambda \mu_1 \dots \mu_{s-1}} - \frac12 \partial_{(\mu_1} \delta \phi'_{\mu_2 \dots \mu_{s-1})} \\
=& \partial^\lambda \left( \partial_\lambda \xi_{\mu_1 \dots \mu_{s-1}}+ \partial_{\mu_1} \xi_{\lambda \mu_2 \dots \mu_{s-1}} + \dots + \partial_{\mu_{s-1}} \xi_{\mu_1 \dots \mu_{s-2}\lambda}\right) - \tfrac12 \, 2 \, \partial^\lambda \partial_{(\mu_1 }  \xi_{\mu_2 \dots \mu_{s-1})\lambda} \\
=& \Box \xi_{\mu_1 \dots \mu_{s-1}} + \partial^\lambda \partial_{(\mu_1 }  \xi_{\mu_2 \dots \mu_{s-1})\lambda} - \partial^\lambda \partial_{(\mu_1 }  \xi_{\mu_2 \dots \mu_{s-1})\lambda} \\
=& \Box \xi_{\mu_1 \dots \mu_{s-1}} \,,
\end{align*}
where to obtain the second equation, we have used $\delta \phi'_{\mu_1 \dots \mu_{s-2}}=2 \partial^\lambda \xi_{\lambda \mu_1 \dots \mu_{s-2}}$ which follows from the tracelessness of the gauge parameter.

%-----------------------------------------
\subsection*{Problem 4}
\noindent Let us first consider the case of $s=3$. We recall that 
\begin{equation}
\mathcal{F}_{\mu \nu \rho} =  F_{\mu \nu \rho} - \frac12 \eta_{(\mu \nu} F'_{\rho)} \,.
\end{equation}
The trace of the Fronsdal tensor is given by
\begin{align}
F'_\mu = \eta^{\nu \rho} F_{\mu \nu \rho} =& \Box \phi'_\mu - \eta^{\nu \rho} \partial^\lambda \left( \partial_\mu \phi_{\nu \rho \lambda} + \partial_\nu  \phi_{\mu \rho \lambda} +  \partial_\rho  \phi_{\mu \nu \lambda}\right) \nonumber\\
& + \eta^{\nu \rho} \left( \partial_\nu \partial_\rho \phi'_\mu + \partial_\nu \partial_\mu \phi'_\rho + \partial_\mu \partial_\rho \phi'_\nu   \right) \nonumber
\\=& 2 \, \Box \phi'_\mu - 2 \, \partial^\lambda \partial^\sigma \phi_{\mu \lambda \sigma} + \partial_\mu \partial^\lambda  \phi'_\lambda \,.
\end{align}
Therefore $\mathcal{F}_{\mu \nu \rho}$ is given by
\begin{align}
\Box \phi_{\mu \nu \rho} - \partial^\lambda \partial_{(\mu} \phi_{\nu \rho)\lambda} + \partial_{(\mu} \partial_{\nu} \phi'_{\rho)} - \frac12 \eta_{(\mu \nu} \left( 2 \, \Box \phi'_{\rho)} - 2 \, \partial^\lambda \partial^\sigma \phi_{\rho) \lambda \sigma} + \partial_{\rho)} \partial^\lambda  \phi'_\lambda \right) \nonumber
\end{align}
In order to prove the symmetry property of the action, we need to show that $\int \psi^{\mu \nu \rho} \mathcal{F}_{\mu \nu \rho}(\phi) = \int \phi^{\mu \nu \rho} \mathcal{F}_{\mu \nu \rho}(\psi)$. The integrand contains terms of the following form:
\begin{itemize}
\item $\psi^{\mu \nu \rho} \Box \phi_{\mu \nu \rho}$ and $\phi'^\mu \Box \phi'_\mu$
\item $\psi^{\mu \nu \rho} \partial_\mu \partial_\lambda  \phi_{\nu \rho}{}^\lambda$ and $\psi'^{\mu } \partial_\mu \partial_\lambda  \phi'^\lambda$  

\end{itemize}
These terms are obviously symmetric upon imposing suitable boundary conditions. The remaining terms are not individually symmetric but combine to
\begin{align}
3 \, \psi^{\mu \nu \rho} \partial_{\mu} \partial_{\nu} \phi'_{\rho} + 3 \, \psi'_{\rho} \partial_\mu \partial_\nu \phi^{\mu \nu \rho} \,.
\end{align}
This combination is again symmetric upon partial integration. Therefore, we have shown the symmetry property of the action for $s=3$.

The case for general $s$ follows along very similar lines. The trace of the Fronsdal tensor is given by
\begin{equation}
F'_{\mu_1 \dots \mu_{s-2}} = 2 \, \Box \phi'_{\mu_1 \dots \mu_{s-2}} + \partial^\sigma \partial_{(\mu_1} \phi'_{\mu_2 \dots \mu_{s-2})\sigma} - 2 \, \partial^{\sigma} \partial^{\lambda} \phi_{\mu_1 \dots \mu_{s-2} \sigma \lambda} \label{eq:traceOfFronsdalTensor}
\end{equation}
We now consider the contraction $\psi^{\mu_1 \dots \mu_s} \mathcal{F}_{\mu_1 \dots \mu_s}(\phi)$. Similarly to the spin-$3$ case, the only non-manifestly symmetric terms are proportional to
\begin{align}
\psi^{\mu_1 \dots \mu_s} \partial_{\mu_1} \partial_{\mu_2} \phi'_{\mu_3 \dots \mu_s} + \psi'_{\mu_1 \dots \mu_{s-2}} \partial_{\mu_{s-1}} \partial_{\mu_s} \phi^{\mu_1 \dots \mu_s} \,.
\end{align}
This sum is again manifestly symmetric.

%-----------------------------------------
\subsection*{Problem 5}
\noindent In order to show this property, it is advantageous to first prove that the Fronsdal tensor is double-traceless. For this we take the trace of \eqref{eq:traceOfFronsdalTensor} which gives
\begin{align}
\eta^{\mu_{s-3} \mu_{s-2}} F'_{\mu_1 \dots \mu_s} = \eta^{\mu_{s-3} \mu_{s-2}} ( 2 \, \Box \phi'_{\mu_1 \dots \mu_{s-2}} + \partial^\sigma \partial_{(\mu_1} \phi'_{\mu_2 \dots \mu_{s-2})\sigma} - 2 \, \partial^{\sigma} \partial^{\lambda} \phi_{\mu_1 \dots \mu_{s-2} \sigma \lambda} ) \nonumber
\end{align}
Since the Fonsdal field is double-traceless, the first term will not contribute. For the same reason, the second term will only contribute if one of the derivatives is carrying either an $\mu_{s-3}$ or $\mu_{s-2}$ index. These terms cancel the contribution from the last term, i.e.
\begin{align}
\eta^{\mu_{s-3} \mu_{s-2}} F'_{\mu_1 \dots \mu_s} =  2 \partial^\sigma \partial^\kappa \phi'_{\mu_1 \dots \mu_{s-4} \sigma \kappa} - 2 \partial^\sigma \partial^\kappa \phi'_{\mu_1 \dots \mu_{s-4} \sigma \kappa} = 0 \,.
\end{align}
Therefore the Fronsdal tensor is double-traceless. We can then easily deduce that
\begin{align}
\eta^{\mu_{s-1} \mu_{s}} \mathcal{F}_{\mu_1 \dots \mu_s} = \eta^{\mu_{s-1} \mu_{s}} ( F_{\mu_1 \dots \mu_s} - \frac12 \, \eta_{( \mu_1 \mu_2} F_{\mu_3 \dots \mu_s)\sigma}{}^\sigma ) \propto F_{\mu_1 \dots \mu_{s-2}}'\,.
\end{align}
The precise proportionality factor is 
\begin{equation}
1-D/2-\tfrac12 2 (s-2) = -\tfrac{D+2s-6}2
\end{equation}
where the last term on the right hand side arises from terms were only either $\mu_{s-1}$ or $\mu_{s}$ is carried by the metric $\eta$ in $\eta_{( \mu_1 \mu_2} F_{\mu_3 \dots \mu_s)}$. There are $2 (s-2)$ such terms. Note that for $s \ge 2$ this factor does not vanish for spacetimes with $D>2$.

\subsection*{Problem 6}
\noindent We want to see that
\begin{align}
\partial^\sigma F_{\mu_1 \dots \mu_{s-1} \sigma} - \frac12 \partial_{(\mu_1} F'_{\mu_2 \dots \mu_{s-1})}  \label{eq:integrandOfGaugeVariationOfFronsdal2} 
\end{align}
vanishes.

The first summand can be straightforwardly evaluated using the definition of the Fronsdal tensor \eqref{eq:fronsdalEquation}:
\begin{align}
&\partial^\sigma \left( \Box \phi_{\mu_1 \dots \mu_{s-1} \sigma} - \partial^\kappa \partial_{(\sigma} \phi_{\mu_1 \dots \mu_{s-1})\kappa} + \partial_{(\sigma} \partial_{\mu_1} \phi'_{\mu_2 \dots \mu_{s-1})} \right) \nonumber \\ =& - \partial^\sigma \partial^\kappa \partial_{(\mu_1} \phi_{\mu_2 \dots \mu_{s-1})\sigma \kappa} + \Box \partial_{(\mu_1} \phi'_{\mu_2 \dots \mu_{s-1})} + \partial_{(\mu_1} \partial_{\mu_2} \partial \cdot \phi'_{\mu_3 \dots \mu_{s-1})} \,. \label{eq:term1}
\end{align}
Using the definition of the Fronsdal tensor \eqref{eq:fronsdalEquation}, the second summand is
\begin{align}
&-\frac12 \partial_{(\mu_1} \left( 2 \, \Box \phi'_{\mu_2 \dots \mu_{s-1})} + \partial^\kappa \partial_{\mu_2} \phi'_{\mu_3 \dots \mu_{s-1})\kappa} -2 \partial^{\kappa} \partial^{\lambda} \phi_{\mu_2 \dots \mu_{s-1})\kappa \lambda}\right) \nonumber
\\=& - \Box \partial_{(\mu_1} \phi'_{\mu_2 \dots \mu_{s-1})} - \partial_{(\mu_1} \partial_{\mu_2} \partial \cdot \phi'_{\mu_3 \dots \mu_{s-1})} + \partial^\sigma \partial^\kappa \partial_{(\mu_1} \phi_{\mu_2 \dots \mu_{s-1})\sigma \kappa} \label{eq:term2}
\end{align}
The last line follows since 
\begin{align}
\partial_{(\mu_1} \left(  \partial^\kappa \partial_{\mu_2} \phi'_{\mu_3 \dots \mu_{s-1}) \kappa} \right) =  2 \, \partial_{(\mu_1} \partial_{\mu_2} \partial \cdot \phi'_{\mu_3 \dots \mu_{s-1})} \,,
\end{align}
Since the symmetrization on the left hand side involves $(s-1)(s-2)$ terms while on the right hand side $\binom{s-1}{2}$ terms are needed.

By comparing \eqref{eq:term2} with \eqref{eq:term1}, we see that \eqref{eq:integrandOfGaugeVariationOfFronsdal2} indeed vanishes. This shows the gauge invariance of the Fronsdal action.

%-----------------------------------------
\subsection*{Problem 7}
\noindent For this problem, we will use the formalism of generating functionals introduced in the solution of Problem 1. Make sure that you have understood this solution before reading further.

In terms of the generating functions, the Fronsdal equation on AdS space \eqref{eq:fronsdalEquationAdS} is given by
\begin{align}
\label{eq:fronsdalEqGenFuncAdS}
F \phi(u) = \left( \Box - u\nabla \, d\nabla + \frac12 \, u\nabla \, u\nabla d^2 - \frac{1}{l^2} \, m_s^2 - \frac{\alpha}{l^2} u^2 \, d^2 \right) \phi(u) = 0 \,.
\end{align}
for $\alpha=1$ and $m_s^2 = s^2 + s(D-6) - 2(D-3)$. In the following, we will indeed show that $\alpha$ and $m_s^2$ are fixed to these values by requiring gauge invariance of \eqref{eq:fronsdalEqGenFuncAdS} under 
\begin{align}
\delta \phi (u) = u \nabla \, \xi(u) \,.
\end{align}
Tracelessness of the gauge parameter enforces $d^2 \, \xi(u) = 0$. Using completely analogous reasoning as in Problem 1, the gauge variation of the Fronsdal tensor can be seen to be be given by
\begin{align*}
F \delta \phi = \left(\Box \, u\nabla - u \nabla [d\nabla, u\nabla] - \frac{2 \alpha}{l^2} \, u^2 \, d\nabla \, - \frac{1}{l^2} \, m_s^2 \, u\nabla \right) \xi(u) \,,
\end{align*}
where we have used the tracelessness of the gauge parameter. One easily can convince oneself that 
\begin{align*}
[d\nabla, u\nabla] = \Box + d^\sigma u^\lambda [\nabla_\sigma, \nabla_\lambda] \,.
\end{align*}
In order to evaluate the last term, we use that
\begin{align}
\label{eq:riemannMaxSym}
R_{\sigma \lambda \tau \rho} = -\frac{1}{l^2} \left( g_{\sigma \tau} g_{\lambda \rho} - g_{\lambda \tau} g_{\sigma \rho}\right) 
\end{align}
for AdS backgrounds and
\begin{align}
\label{eq:commutatorCovDerivGenFunc}
 [\nabla_\sigma, \nabla_\lambda] f(u) = R_{\sigma \lambda \tau \rho} \, u^\tau d^\rho \, f(u) \,,
\end{align}
for an arbitrary function $f$.
Using these identities and tracelessness of the gauge parameter, one obtains after some straightforward algebra
\begin{align}
F \delta \phi(u) = \left\{ [\Box,u\nabla] + \frac{1}{l^2} \left( u\nabla (D + ud) ud - 2 \, u\nabla \, ud \right) - \frac{2 \alpha}{l^2} \, u^2 \, d\nabla \, - \frac{1}{l^2} \, m_s^2 \, u\nabla \right\} \xi(u) \,.
\end{align}
We will now use the identity
\begin{align}
\label{eq:magicIdentity}
[u \nabla, \Box] = \frac1{l^2} \left( u\nabla \, (2 ud + D -1 ) - 2 \, u^2 \,d\nabla \right) \,,
\end{align}
which we will prove later. By applying this identity, we arrive at
\begin{align*}
F \delta \phi(u) = \frac{-1}{l^2} \left\{\left( m_s^2 - [s^2 + s(D-6) - 2(D-3)] \right) u \nabla - 2(\alpha - 1) \, d\nabla \right\} \xi(u) \,
\end{align*}
where we have used $ud \, \xi(u) = (s-1) \xi(u)$. This fixes the mass to be $m_s^2 = s^2 + s(D-6) - 2(D-3)$ and the constant $\alpha=1$.

The only step left to complete the proof, is to derive identity \eqref{eq:magicIdentity}. In order to do so, it is useful to first consider 
\begin{align*}
[u\nabla, \nabla_\sigma] \nabla^\sigma \, f(u) &= u^\lambda \, R_{\lambda \sigma}{}^\sigma{}_{\tau} \, \nabla^\tau f(u) + u^\lambda \, \nabla^\sigma R_{\lambda \sigma \tau \rho} \, u^\tau d^\rho \, f(u) \\
&= -\frac{1}{l^2} \left( u \nabla (ud + D -1 ) - u^2 d\nabla \right) f(u) \,,
\end{align*}
where we used covariant constancy of the Riemann tensor for AdS spacetime to obtain the first equation and \eqref{eq:commutatorCovDerivGenFunc} along with \eqref{eq:riemannMaxSym} for the last line. From this, the identity \eqref{eq:magicIdentity} follows straightforwardly by again using \eqref{eq:commutatorCovDerivGenFunc} and \eqref{eq:riemannMaxSym}.

%-----------------------------------------
\subsection*{Problem 8}
\noindent In the following, we again use the method of generating functions introduced in the solutions to Problem 1.

The left hand side of the modified Bianchi identity \eqref{eq:modifiedBinachiIdent} can in this language be rewritten as
\begin{align}
d \partial F(u) - \frac12 \, u\partial \, d^2 F(u) \,.
\end{align}
We now plug in the definition of $F(u)$ given by \eqref{eq:fronsdalGenFunc} and sort by powers of the $d$ operator. This leads to
\begin{align}
\left \{d\partial \Box - d\partial \, u\partial \, d\partial - \frac12 \, u\partial \, d^2 \Box + \frac12 \, d\partial \, (u\partial)^2 \, d^2 + \frac12 \, u \partial \, d^2 \, u\partial \, d\partial - \frac14 \, u\partial \, d^2 \, (u\partial)^2 \, d^2 \right\} \phi (u)
\end{align}
Using the identities $[d^2, (u\partial)^2]=4 \, d\partial \, u \partial - 2 \Box$ and 
\begin{align*}
\frac12 \left( d\partial \, (u\partial)^2 \, d^2 + u \partial \, d^2 \, u\partial \, d\partial \right) = u\partial \, d\partial \, u\partial \, d^2 + u \partial \, (d\partial)^2
\end{align*}
We arrive at 
\begin{align*}
\left\{d\partial \Box - d\partial \, u \partial \, d\partial + u\partial \, (d\partial)^2 - \frac14 (u\partial)^3 \, d^4 \right\} \phi(u)
\end{align*}
which using $[u\partial,d\partial]=-\Box$ in the second summand is equal to
\begin{align}
-\frac14 (u\partial)^3 \,d^4 \phi(u)
\end{align}
We therefore conclude that the Bianchi identity is indeed violated by a double-trace term. The factor can be easily determined by plugging in the definition of $\phi(u)$:
\begin{align*}
-\frac14 \, \frac1{s!} \, s(s-1)(s-2) \, \partial_{\mu_1} \partial_{\mu_2} \partial_{\mu_3} \phi''_{\mu_4 \dots \mu_s} \, u^{\mu_1} \dots u^{\mu_s}
= - \frac32 \, \frac1{s!} \, \partial_{(\mu_1} \partial_{\mu_2} \partial_{\mu_3} \phi''_{\mu_4 \dots \mu_s)}  \, u^{\mu_1} \dots u^{\mu_s} \,.
\end{align*}

\bibliography{refs}

\providecommand{\href}[2]{#2}\begingroup\raggedright\begin{thebibliography}{10}

\bibitem{Rahman:2015pzl}
R.~Rahman and M.~Taronna, \emph{{From Higher Spins to Strings: A Primer}},
  \href{https://arxiv.org/abs/1512.07932}{{\tt 1512.07932}}.

\bibitem{Didenko:2014dwa}
V.~E. Didenko and E.~D. Skvortsov, \emph{{Elements of Vasiliev theory}},
  \href{https://arxiv.org/abs/1401.2975}{{\tt 1401.2975}}.

\bibitem{Campoleoni:2010zq}
A.~Campoleoni, S.~Fredenhagen, S.~Pfenninger and S.~Theisen, \emph{{Asymptotic
  symmetries of three-dimensional gravity coupled to higher-spin fields}},
  \href{http://dx.doi.org/10.1007/JHEP11(2010)007}{\emph{JHEP} {\bf 11} (2010)
  007}, [\href{https://arxiv.org/abs/1008.4744}{{\tt 1008.4744}}].

\bibitem{Fronsdal:1978rb}
C.~Fronsdal, \emph{{Massless Fields with Integer Spin}},
  \href{http://dx.doi.org/10.1103/PhysRevD.18.3624}{\emph{Phys. Rev.} {\bf D18}
  (1978) 3624}.

\bibitem{Weinberg:1995mt}
S.~Weinberg, \emph{{The Quantum theory of fields. Vol. 1: Foundations}}.
\newblock Cambridge University Press, 2005.

\bibitem{Bekaert:2006py}
X.~Bekaert and N.~Boulanger, \emph{{The Unitary representations of the Poincare
  group in any spacetime dimension}},  in \emph{{2nd Modave Summer School in
  Theoretical Physics Modave, Belgium, August 6-12, 2006}}, 2006.
\newblock \href{https://arxiv.org/abs/hep-th/0611263}{{\tt hep-th/0611263}}.

\bibitem{Prokushkin:1998bq}
S.~F. Prokushkin and M.~A. Vasiliev, \emph{{Higher spin gauge interactions for
  massive matter fields in 3-D AdS space-time}},
  \href{http://dx.doi.org/10.1016/S0550-3213(98)00839-6}{\emph{Nucl. Phys.}
  {\bf B545} (1999) 385}, [\href{https://arxiv.org/abs/hep-th/9806236}{{\tt
  hep-th/9806236}}].

\bibitem{Vasiliev:1999ba}
M.~A. Vasiliev, \emph{{Higher spin gauge theories: Star product and AdS
  space}},  \href{https://arxiv.org/abs/hep-th/9910096}{{\tt hep-th/9910096}}.

\bibitem{Klebanov:2002ja}
I.~R. Klebanov and A.~M. Polyakov, \emph{{AdS dual of the critical O(N) vector
  model}}, \href{http://dx.doi.org/10.1016/S0370-2693(02)02980-5}{\emph{Phys.
  Lett.} {\bf B550} (2002) 213--219},
  [\href{https://arxiv.org/abs/hep-th/0210114}{{\tt hep-th/0210114}}].

\bibitem{Sezgin:2003pt}
E.~Sezgin and P.~Sundell, \emph{{Holography in 4D (super) higher spin theories
  and a test via cubic scalar couplings}},
  \href{http://dx.doi.org/10.1088/1126-6708/2005/07/044}{\emph{JHEP} {\bf 07}
  (2005) 044}, [\href{https://arxiv.org/abs/hep-th/0305040}{{\tt
  hep-th/0305040}}].

\bibitem{Gaberdiel:2012uj}
M.~R. Gaberdiel and R.~Gopakumar, \emph{{Minimal Model Holography}},
  \href{http://dx.doi.org/10.1088/1751-8113/46/21/214002}{\emph{J. Phys.} {\bf
  A46} (2013) 214002}, [\href{https://arxiv.org/abs/1207.6697}{{\tt
  1207.6697}}].

\bibitem{Giombi:2016ejx}
S.~Giombi, \emph{{TASI Lectures on the Higher Spin - CFT duality}},
  \href{https://arxiv.org/abs/1607.02967}{{\tt 1607.02967}}.

\bibitem{Chang:2012kt}
C.-M. Chang, S.~Minwalla, T.~Sharma and X.~Yin, \emph{{ABJ Triality: from
  Higher Spin Fields to Strings}},
  \href{http://dx.doi.org/10.1088/1751-8113/46/21/214009}{\emph{J. Phys.} {\bf
  A46} (2013) 214009}, [\href{https://arxiv.org/abs/1207.4485}{{\tt
  1207.4485}}].

\bibitem{Gaberdiel:2014cha}
M.~R. Gaberdiel and R.~Gopakumar, \emph{{Higher Spins \& Strings}},
  \href{http://dx.doi.org/10.1007/JHEP11(2014)044}{\emph{JHEP} {\bf 11} (2014)
  044}, [\href{https://arxiv.org/abs/1406.6103}{{\tt 1406.6103}}].

\bibitem{Gaberdiel:2015wpo}
M.~R. Gaberdiel and R.~Gopakumar, \emph{{String Theory as a Higher Spin
  Theory}}, \href{http://dx.doi.org/10.1007/JHEP09(2016)085}{\emph{JHEP} {\bf
  09} (2016) 085}, [\href{https://arxiv.org/abs/1512.07237}{{\tt 1512.07237}}].

\bibitem{Francia:2010ap}
D.~Francia, \emph{{On the Relation between Local and Geometric Lagrangians for
  Higher spins}},
  \href{http://dx.doi.org/10.1088/1742-6596/222/1/012002}{\emph{J. Phys. Conf.
  Ser.} {\bf 222} (2010) 012002}, [\href{https://arxiv.org/abs/1001.3854}{{\tt
  1001.3854}}].

\bibitem{Francia:2005bu}
D.~Francia and A.~Sagnotti, \emph{{Minimal local Lagrangians for higher-spin
  geometry}},
  \href{http://dx.doi.org/10.1016/j.physletb.2005.08.002}{\emph{Phys. Lett.}
  {\bf B624} (2005) 93--104}, [\href{https://arxiv.org/abs/hep-th/0507144}{{\tt
  hep-th/0507144}}].

\end{thebibliography}\endgroup
\bibliographystyle{JHEP}

\end{document}